\documentclass[twocolumn,superscriptaddress,prl,10pt]{revtex4-1}

\usepackage{amsmath}
\usepackage{amsfonts}
\usepackage{color}
\usepackage[colorlinks,bookmarks=false,citecolor=blue,linkcolor=red,urlcolor=blue]{hyperref}
\usepackage{ulem}

\usepackage{graphicx,subfigure}

\def \be {\begin{equation}}
\def \ee {\end{equation}}
\def \ba {\begin{array}}
\def \ea {\end{array}}
\def \bea {\begin{eqnarray}}
\def \eea {\end{eqnarray}}

\def \ble {\begin{widetext}\begin{equation}}
\def \ele {\end{equation}\end{widetext}}
\def \blea {\begin{widetext}\begin{eqnarray}}
\def \elea {\end{eqnarray}\end{widetext}}

\def \nn {\nonumber}

\def \a {\alpha}

\def \G {\Gamma}

\def \D {\Delta}
\def \e {\epsilon}
\def \ve {\varepsilon}
\def \m {\mu}

\def \l {\lambda}
\def \L {\Lambda}
\def \s {\sigma}

\def \r {\rho}

\def \mD {\mathcal D}

\def \mS {\mathcal S}

\def \mX {\mathcal X}
\def \mY {\mathcal Y}

\def \cT {{\mathcal T}}

\def \cX {{\mathcal X}}

\def \p {\partial}
\def \f {\frac}

\def \sr {\sqrt}
\def \td {\tilde}

\def \inf {\infty}

\def \lag {\langle}
\def \rag {\rangle}

\def \ep {\mathrm{e}}
\def \ii {\mathrm{i}}

\def \tr {\mathrm{tr}}

\def \and {{\mathrm{and}}}

\begin{document}

\title{Subsystem Trace Distance in Quantum Field Theory}

\author{Jiaju Zhang}
\affiliation{SISSA and INFN, Via Bonomea 265, 34136 Trieste, Italy}

\author{Paola Ruggiero}
\affiliation{SISSA and INFN, Via Bonomea 265, 34136 Trieste, Italy}

\author{Pasquale Calabrese}
\affiliation{SISSA and INFN, Via Bonomea 265, 34136 Trieste, Italy}
\affiliation{International Centre for Theoretical Physics (ICTP), Strada Costiera 11, 34151 Trieste, Italy}

\begin{abstract}
We develop a systematic method to calculate the trace distance between two reduced density matrices in 1+1 dimensional quantum field theories. 
The approach exploits the path integral representation of the reduced density matrices and an ad hoc replica trick. 
We then extensively apply this method to the calculation of the distance between reduced density matrices of one interval of length $\ell$ 
in eigenstates of conformal field theories. 
When the interval is short, using the operator product expansion of twist operators, we obtain a universal form for the leading order in $\ell$ of the trace distance. 
We compute the trace distances among the reduced density matrices of several low lying states in two-dimensional free 
massless boson and fermion theories. 
We compare our analytic conformal results with numerical calculations in XX and Ising spin chains finding perfect agreement.
\end{abstract}

\maketitle



During recent times, in several disconnected fields of physics emerged the necessity to 
characterise the properties of extended subsystems rather than of the entire system. 
A first important example is the non-equilibrium dynamics of isolated quantum systems:
while the whole system remains in a pure state, 
subsystems are described by statistical ensembles for large times \cite{bs-08,gogolin-2015,calabrese-2016,vidmar-2016,ef-16,dls-13,kaufman-2016,ac-17}.
Another subject where the physics of the subsystem matters is the black hole information loss paradox \cite{h-74} and its relation to the distinguishability of states 
in gauge field theories through the gauge/gravity duality \cite{Maldacena:1997re,rt-06,r-10,m1,ms-13,Fitzpatrick:2016ive}.
Indeed, the large interest in the physics of subsystems was the main reason of the vast theoretical \cite{intro1,intro2,intro3} and 
experimental \cite{exp1,kaufman-2016,evdcz-18,lukin-18} activity aimed to characterise the entanglement 
of extended quantum systems.  
Yet, it is  very important not only to have information about the subsystems, but also to have a notion of distance between the subsystems' configurations, 
more precisely between the reduced density matrices (RDM) of a subsystem in two different states. 

%

Several different measures  of the distance between density matrices exist and are widely used in quantum information \cite{nc-book,watrous-book}.
For example, given two density matrices $\rho$ and $\sigma$ acting on the same Hilbert space, 
a family of distances, depending on a continuous parameter $n$, is provided by the $n$-distances
\be \label{Dnrs}
D_n(\r,\s) = \f{1}{2^{1/n}} \| \r - \s \|_n,
\ee
where the $n$-norm is $\|\L\|_n = ( \sum_i \l_i^n )^{1/n}$ 
with $\l_i$ being the singular values of $\L$, i.e., the eigenvalues of $\sr{\L^\dag \L}$. When $\L$ is Hermitian, $\l_i$ are just the absolute values of the eigenvalues of $\L$.
(The normalization $2^{1/n}$  ensures $0 \leq D_n(\r,\s) \leq 1$.)
%
In finite dimensional Hilbert spaces all norms are equivalent (in the sense that they bound each other), 
but this ceases to be the case for infinite dimensional spaces and we are interested 
in quantum field theories (QFT). 
Furthermore, even in finite dimensions, the bounds between norms depend on the dimension and
so the distances are not on equal footing when comparing subsystems of different sizes.
There are several reasons why the trace distance $D_1$ is special, e.g. 
the difference of expectation values of an operator $O$ in different states is bounded 
as $|\tr (\r-\s) O|\leq D_1(\r,\s) ||O||_1$, and the bound has no factor depending on the Hilbert space dimension, as it would be the case 
for other norms, see e.g. \cite{fe-13b}.

It is however extremely difficult to evaluate analytically $D_1$: even for Gaussian states there is no way to compute it from the two-point correlators
 (see e.g. \cite{fe-13b}).
This is indeed one of the reasons why for quantum field theories there has been a huge activity \cite{hol-rel-entropy,lashkari2014,bhm-14,lashkari2016,rc-17,su-16,su-17,casini-2016,Nakagawa:2017fzo,mrc-18}
 in quantifying the relative entropy
$S(\r\|\s) = \tr(\r\log\r) - \tr(\r\log\s)$, which provides an idea about the distance of $\r$ and $\s$, but it is not even a metric since it is not symmetrical.

In this Letter we develop a systematic method to calculate the trace distance between two RDMs in quantum field theories.
This is based on an ad hoc replica trick: we first calculate the $n$-distance \eqref{Dnrs} for a general {\it even} integer $n$, we analytically continue it to 
real $n$, and finally we take the limit $n \to 1$ to get the trace distance. 
(This trick is reminiscent of the one for entanglement negativity in QFT \cite{cct-12}.)
For even $n$, we express $D_n$ as a correlation function in a replicated worldsheet, that, in 1+1 dimensions, can be written in 
terms of the  twist operators of Refs. \cite{cc-04,cc-09,ccd-08}.
We will then exploit this method to provide very general results for the distance between primary states in 2D conformal field theories (CFT)
and use the operator product expansion (OPE) of the twist fields \cite{headrick,cct-11,gr-12,cz-13,cwz-16,lwz-16,hlz-17} 
for short intervals. 
In this Letter we just introduce our replica approach and show how it works with a few examples. 
Most details are left to a technical forthcoming publication \cite{work-td}.

\paragraph{{\it General description in QFT.}}

\begin{figure}[t]
  \centering
  \includegraphics[width=0.48\textwidth]{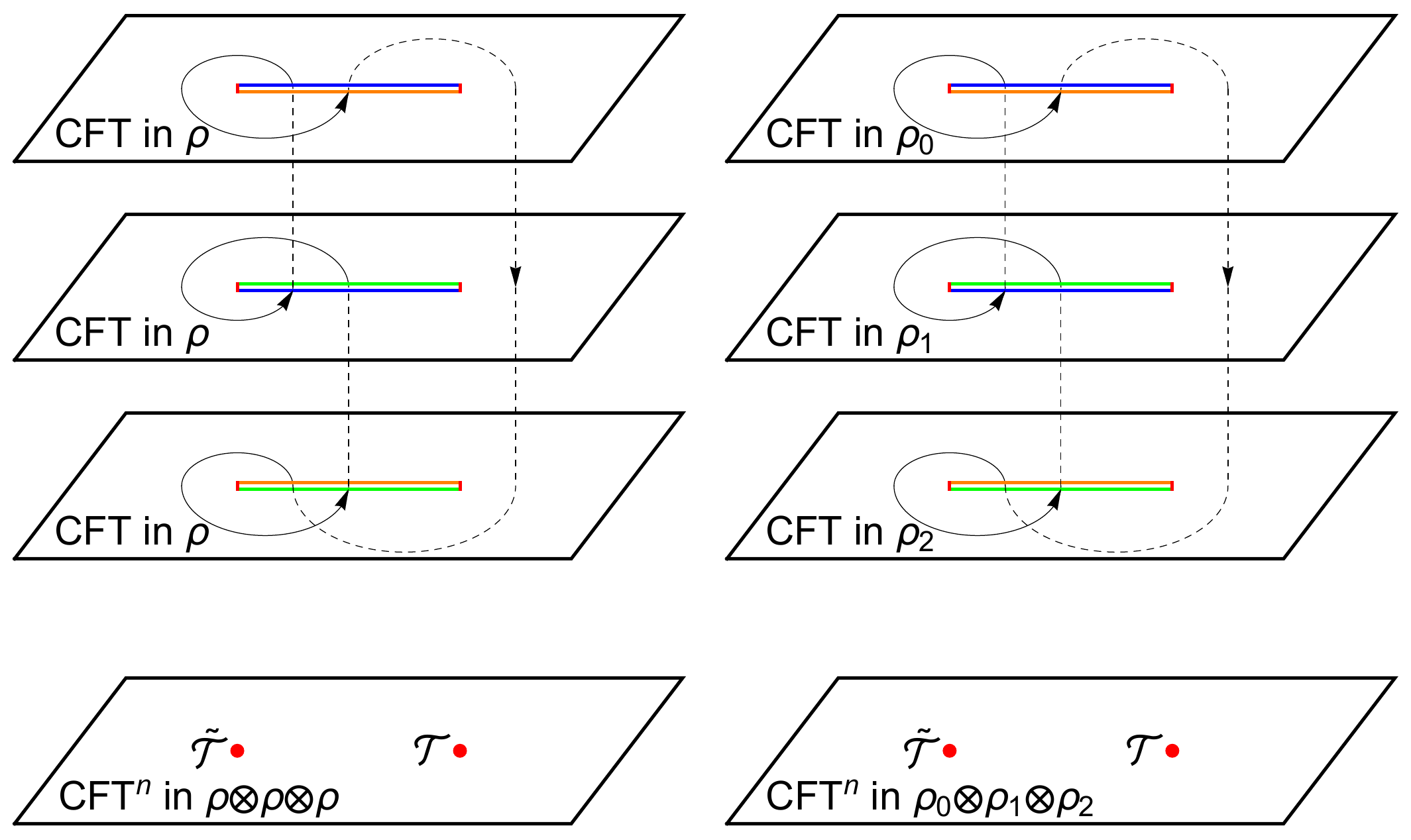}\\
  \caption{The replica trick to calculate $\tr \r_A^n$ (left) and for each term in the sum in the r.h.s. of \eqref{expansion}, 
  $\tr (\r_{A,0}\r_{A,1}\cdots\r_{A,n-1})$ (right). 
  Top: path integral  in terms of Riemann surfaces. Bottom: Equivalent representation in the $n$-fold theory, Eqs.~\eqref{trArAn} and \eqref{tr-rel}.
 We show the case $n=3$ as an example. }\label{replica}
\end{figure}

We consider a generic QFT in a state described by the density matrix $\r$ and we focus on a spatial subsystem $A$. 
The reduced density matrix $\r_A = \tr_{\bar A}\r$ is obtained by tracing out the degrees of freedom of $\bar A$,  the complement of $A$. 
The bipartite entanglement between $A$ and $\bar A$ may be measured by the celebrated entanglement entropy $S_A = - \tr (\r_A\log\r_A) $ which can be obtained 
from the replica limit of $\tr \r_A^n$ 
\cite{cc-09,cc-04}.
In the path integral Euclidean formalism and for $n \in \mathbb{N}$, such traces correspond to partition functions on an $n$-sheeted Riemann surface (see Fig.~\ref{replica}, left), with the sheet $j$ representing $\rho_{A,j}$ (the $j$-th copy of the state $\rho_A$ of the initial QFT). 
Moreover, for one interval embedded in a 1D systems (equivalently 2D QFT), the same quantities can be interpreted as two-point functions 
(or, more generally, $2m$-point functions if the subsystem consists of $m$ disjoint intervals) 
evaluated in the state $\r_n = \otimes_{j=0}^{n-1} \rho_{A,j}$ of the corresponding $n$-fold theory \cite{cc-04,ccd-08} 
\be \label{trArAn}
\tr \r_A^n = \lag \cT(\ell,\ell)\td\cT(0,0) \rag_{\r_n}.
\ee
$\cT$ and $\td\cT$ are called twist and antitwist operators respectively. This form is particularly useful in the context of CFT, where $\cT$ and $\td\cT$ are primary operators of the $n$-fold theory (dubbed CFT$^n$) with conformal weights $h_n = \bar h_n = \f{c(n^2-1)}{24n}$  \cite{cc-04}. Here $c$ is the central charge of the 
single-copy CFT. For an interval of length $\ell$ in an infinite system, this leads to the famous result 
$ \tr \r_A^n = c_n ({\ell}/{\e})^{-4h_n}$ where $c_n$ is the normalization constant \cite{cc-09} ($c_1=1$) of the twist operators and $\e$ the UV cutoff.

Here, we are interested in the trace distance (i.e. \eqref{Dnrs} for $n=1$) between two generic RDMs $\rho_A, \sigma_A$. 
For an arbitrary {\it even} integer $n_e$ we have $\|\r_A-\s_A\|_{n_e}^{n_e}=\tr (\r_A-\s_A)^{n_e}$. Therefore, if we are able to compute $\tr (\r_A-\s_A)^{n_e}$ and analytically continue it as a function of $n_e$ from even integers to generic real values, then the trace distance is obtained via the following {\it replica trick}
\be
\label{eq:replica}
D_{1} (\r_A, \s_A)= \f12\lim_{n_e \to 1} \tr (\r_A-\s_A)^{n_e}.
\ee
We stress that for odd $n_o$ instead, $\tr (\r_A-\s_A)^{n_o}$ does {\it not} provide the $n_o$-distance. 
Hence the limit $n_o\to 1$ just gives the trivial result $\tr(\r_A-\s_A)=0$, in full analogy to the negativity replica trick \cite{cct-12}.

To proceed, we note that $\tr (\r_A-\s_A)^{n}$ may be expanded as 
\be \label{expansion}
\tr (\r_A-\s_A)^{n} = \sum_{\mathcal{S}} (-)^{|\mathcal{S}|} \tr \left( \rho_{0_{\mathcal{S}}} \cdots \rho_{(n-1)_{\mathcal{S}}} \right)  ,
\ee
where the summation $\mathcal{S}$ is over all the subsets of $\mathcal{S}_0 = \{ 0, \cdots, n-1 \}$, $| \mathcal{S}|$ is the cardinality of $\mathcal{S}$ and $\rho_{j_{\mathcal{S}}} =  \sigma_A $ if $j \in \mathcal{S}$ and $\rho_A$ otherwise. Crucially, each term in the sum appearing in the r.h.s. of Eq.~\eqref{expansion} is related to a partition function on an
 $n$-sheeted Riemann surface (see Fig. \ref{replica}, right) and, again, in a 2D QFT, it can be written as a two-point  function of twist fields (cfr., e.g., \cite{dei-18})
\be
\label{tr-rel}
\tr \left(  \rho_{0_{\mathcal{S}}} \cdots \rho_{{(n-1)}_{\mathcal{S}}}  \right) = \lag \cT(\ell,\ell)\td\cT(0,0) \rag_{\otimes_j \rho_{j_{ \mathcal{S} }}}.
\ee
Such objects are the same appearing in the replica trick for the relative entropy \cite{lashkari2014,lashkari2016} and in some cases have been explicitly computed 
\cite{lashkari2014,lashkari2016,su-16,su-17,rc-17}. Still, performing the sum in Eq.~\eqref{expansion} and obtaining its analytic continuation is not  an easy task.

\paragraph{Trace distance in 2D CFT.} We now  consider a 1D system whose scaling limit is described by a 2D CFT. 
We focus on periodic systems of total length $L$ and on a subsystem being an interval of length $\ell$ (say $A=[0,\ell]$). 
In Euclidean path integral, the CFT is described by the complex coordinate $z= x + \ii t$ (and its conjugate $\bar{z}$), with $x \in [0, L]$ and $t \in \mathbb{R}$, i.e. 
the worldsheet is an infinite cylinder.
We consider the distance between RDMs of orthogonal eigenstates; as we shall see while the distance between the entire states is maximal, subsystems may be rather close. 
For a general primary operator $\mX$ of conformal weights $(h_{\mX}, \bar{h}_{\mX})$, scaling dimension $\Delta_{\mX} = h_{\mX} + \bar{h}_{\mX}$ and spin $s_{\mX} =h_{\mX}- \bar{h}_{\mX}$, we can use the state-operator correspondence to construct the ket $| \mX \rangle $ and bra $\langle \mX |$ states.
The associated density matrix restricted to $A$ is simply  $\r_{\mX} = \tr_{\bar{A} }|\mX\rag\lag\mX|$.
We exploit Eq.~\eqref{expansion} to compute $\tr \left( \r_{\mX} - \r_{\mY} \right)^n$ for two RDMs associated to two primary operators $\mX$ and $\mY$. 
For such states, each term of the sum in the r.h.s. corresponds to a $2n$-point correlation function of the fields $\mX$ and $\mY$ on the Riemann surface \cite{sierra,berganza},
which is mapped to the complex plane $\mathbb{C}$ through the  map
\be
f(z)= \left( \frac{z - e^{ 2 \pi \ii \ell/L}}{z-1} \right)^{1/n}.
\ee
The final result for the entire sum in Eq.~\eqref{expansion} can be written as a sum of such correlation functions and reads
\begin{multline}
 \label{trArXrY}
 \tr( \r_\mX - \r_\mY )^n = c_n \Big( \f{L}{\pi\e}\sin\f{\pi\ell}{L} \Big)^{-4h_n} \\
                              \sum_{\mS} \Big\{
                             (-)^{|\mS|}
                             \ii^{2( |\bar\mS|s_\mX + |\mS|s_\mY )}
                             \Big( \f{2}{n}\sin\f{\pi\ell}{L} \Big)^{2(|\bar\mS|\D_\mX+|\mS|\D_\mY)} \\
\times \Big\lag
   \Big[ \prod_{j \in \bar \mS} \Big( f_{j,\ell}^{h_\mX} \bar f_{j,\ell}^{\bar h_\mX} f_j^{h_\mX} \bar f_j^{\bar h_\mX}  \mX(f_{j,\ell},\bar f_{j,\ell}) \mX^\dag(f_{j},\bar f_{j}) \Big)\Big] \\
\times \Big[ \prod_{j \in \mS} \Big( f_{j,\ell}^{h_\mY} \bar f_{j,\ell}^{\bar h_\mY} f_j^{h_\mY} \bar f_j^{\bar h_\mY} \mY(f_{j,\ell},\bar f_{j,\ell}) \mY^\dag(f_{j},\bar f_{j}) \Big)\Big]
   \Big\rag_{\mathbb{C}} \Big\}.
\end{multline}
Here $\bar\mS = \mS_0/\mS$, $f_j = \ep^{\f{2\pi\ii j}{n}} $ and $ f_{j,\ell} = \ep^{\f{2\pi\ii}{n}(j+\f{\ell}{L})}$.
Note that in the limit $n \to 1$ the dependence on the ultraviolet cutoff $\epsilon$ washes out. Importantly, this means that the trace distance is {\it cutoff independent}, 
scale invariant (i.e., it depends on $\ell$ and $L$ only through $\ell/L$), and universal.
We can also introduce scale-invariant cutoff-independent quantities for the $n$-distance as
\be
\label{mFn} 
\mD_{n} (\r,\s) = \frac{\tr \left( \r - \s \right)^{n}}{\tr \r^{n}},
\ee
with replica limit from~\eqref{eq:replica} 
\be
\label{eq:replica2}
D_{1} (\r, \s) = \frac{1}{2} \lim_{n_e \to 1} \mD_{n_e} (\r, \s).
\ee

\paragraph{Small interval expansion.}
When the interval is short $\ell \ll L$, one can use the OPE of twist operators \cite{headrick,cct-11,gr-12,cz-13} to expand the partition function (\ref{trArAn}) as a sum of 
 one-point functions in CFT$^n$.
When the state $\r$ is translational invariant, in the OPE of twist operators in (\ref{trArAn}) we only need to include the CFT$^n$ quasiprimary operators that are direct products of the nonidentity quasiprimary operators $\{\cX\}$ of the original CFT, i.e. operators of the form $\mX_1^{j_1}\cdots\mX_k^{j_k}$ (see \cite{cwz-16,rtc-18}  for a discussion).
For example $ \tr \r_A^n$ is expanded as \cite{cwz-16,lwz-16,hlz-17} 
\begin{multline}
 \tr \r_A^n =
c_n \Big(\f{\ell}{\e}\Big)^{-4h_n}
\Big[ 1+
\sum_{k=1}^n
\sum_{\{\mX_1,\cdots,\mX_k\}}
\ell^{\D_{\mX_1}+\cdots+\D_{\mX_n}} \\
\times
b_{\mX_1\cdots\mX_k}
\lag \mX_1 \rag_\r
\cdots
\lag \mX_k \rag_\r  \Big].
\label{expR}
\end{multline}
The coefficients $b_{\mX_1\cdots\mX_k}$ have been defined in \cite{cwz-16} and they are related to the OPE coefficients of the CFT$^n$ operators $\mX_1^{j_1}\cdots\mX_k^{j_k}$.

Similarly, for the RDMs $\r_A$, $\s_A$ of two translationally invariant states $\r$, $\s$, following Ref. \cite{hlz-17}, one gets 
\bea \label{trArAmsAn1}
&& \tr (\r_A-\s_A)^n = c_n \Big(\f{\ell}{\e}\Big)^{-4h_n}
\sum_{\{\mX_1,\cdots,\mX_n\}} \big[
 \ell^{\D_{\mX_1}+\cdots+\D_{\mX_k}}  \nn\\
&& ~
\times b_{\mX_1\cdots\mX_n}
\big(\lag \mX_1 \rag_\r - \lag \mX_1 \rag_\s\big)
\cdots
\big(\lag \mX_n \rag_\r - \lag \mX_n \rag_\s\big) \big].
\eea
Given that the two states $\r$, $\s$ are different, quasiprimary operators $\phi$ such that  
\be
\label{condition}
\lag\phi\rag_\r - \lag\phi\rag_\s \neq 0,
\ee
should exist. 
The OPE in \eqref{trArAmsAn1} is dominated by the quasiprimary operator $\phi$ satisfying \eqref{condition}
with the smallest scaling dimension $\D_\phi$. 
For simplicity, we assume that there is only a single $\phi$ with these properties (but this condition can be relaxed). 
The operator $\phi$ is normalized as $\langle \phi(0,0)\phi(z,\bar z)\rangle_{\mathbb C}= z^{-2h_\phi}\bar z^{-2\bar h_\phi}$ and
is a bosonic operator, i.e.,  its spin $s_\phi$ is an integer.
Hence, for general even integer $n_e$, from (\ref{trArAmsAn1}) we get 
\be \label{trArAmsAn2}
\frac{\tr (\r_A-\s_A)^{n_e}}{ \tr \r_A^{n_e}} = 
                  \ell^{n_e \D_\phi} b_{\phi\cdots\phi} \big(\lag \phi \rag_\r - \lag \phi \rag_\s\big)^{n_e} 
                          +o(\ell^{n_e \D_\phi}), 
\ee
with $\phi\cdots\phi$ denoting the direct product of $n_e$ $\phi$'s.
In Eqs. \eqref{expR}, \eqref{trArAmsAn1}, and \eqref{trArAmsAn2}, the $L$ dependence  is hidden in $\lag \mX_k \rag_\r\propto L^{-\Delta_k}$, making each addendum a
function only of $\ell/L$.

For arbitrary $n_e$, (\ref{trArAmsAn1}) leads to the expansion of the $n_e$-distance $D_{n_e} (\r_A,\s_A)$ and (\ref{trArAmsAn2}) gives its leading order for a short interval.
The replica limit $n_e \to1$ of (\ref{trArAmsAn2}) provides a universal expression for the leading order trace distance in short interval expansion
\be \label{DrAsA}
D_1(\r_A,\s_A) = \f{x_\phi | {\lag\phi\rag_\r - \lag\phi\rag_\s}| }{2} \ell^{\D_\phi}
               + o(\ell^{\D_\phi}),
\ee
with  ($n_e = 2 p$, $ p \in \mathbb{N}$) 
\be \label{xphi}
x_\phi = \lim_{p \to 1/2} \f{\ii^{2 p s_\phi}}{(2p)^{2p\D_\phi}}
                          \Big\lag \prod_{j=0}^{2p-1}
                          \big[ \ep^{\f{\pi\ii j}{p}s_\phi} \phi(\ep^{\f{\pi\ii j}{p}},\ep^{-\f{\pi\ii j}{p}}) \big] \Big\rag_{\mathbb{C}}.
\ee
This short distance result is remarkable, although it only applies to the case with no degeneracy at scaling dimension $\D_\phi$, i.e., that at scaling dimension $\D_\phi$ there only exists one quasiprimary operator $\phi$ with $\lag\phi\rag_\r - \lag\phi\rag_\s \neq 0$.
The correlation function appearing in \eqref{xphi} has been explicitly calculated for several operators in \cite{sierra,berganza} 
and analytically continued in \cite{elc-13}.


Finally, we also get an upper bound for $x_\phi$
\be \label{bound}
x_\phi \leq x_{\rm{max}}(\D_\phi) = \sr{\f{\pi^{1/2} \G(\D_\phi+1)}{2^{2\D_\phi+1}\G(\D_\phi+\f32)}},
\ee
which depends solely on the scaling dimension. 
Eq.~\eqref{bound} follows from the  Pinsker's inequality $D(\r,\s) \leq \sr{S(\r\|\s) /2}$
and known results for the (universal) leading order of the relative entropy \cite{su-16,su-17,hlz-17}.

\paragraph{2D free massless boson theory.}

We consider the 2D free massless boson theory compactified on  circle of unit radius, corresponding to a 2D CFT with central charge $c=1$. 
We denote by $\varphi$, $\bar\varphi$ 
the holomorphic and anti-holomorphic parts of the scalar, respectively. The primary operators are
the currents $J=\ii\p\varphi$, $\bar J=\ii\bar\p\bar\varphi$ with conformal weights $(1,0)$, $(0,1)$
as well as the vertex operators $V_{\a,\bar\a}=\exp(\ii\a\varphi+\ii\bar\a\bar\varphi)$ with $\a,\bar\a=0,\pm1,\cdots$ and conformal weights $({\a^2}/{2},{\bar\a^2}/{2})$.
The low energy excited states are obtained by acting on the vacuum with such operators.
We denote the associated RDMs as $\r_{\a,\bar\a}$, $\r_{J}$, $\r_{\bar J}$ and $\r_{0,0}=\r_0$. 
Details about the 2D free massless boson theory can be found in \cite{DiFrancesco:1997nk,Blumenhagen:2009zz}.

We start by considering Eq.~\eqref{mFn} for two generic vertex operators when it can be written as
\begin{multline}
\label{mFn-vertex}
 \mD_n [\Delta \alpha] \equiv \mD_n (\rho_{\alpha, \bar{\alpha}}, \rho_{ \alpha' , \bar{\alpha}' })= \\
  \sum_{k=0}^n (-)^k \sum_{0 \leq j_1 < \cdots < j_k \leq n-1} 
   d_n ( \{ j_1, \cdots, j_k \})^{ \Delta \alpha}     ,
\end{multline}
where $\Delta \alpha \equiv (\a-\a')^2+(\bar\a-\bar\a')^2$ and
\begin{multline}
\label{FS}
d_n(\mS) = \Big( \f{\sin{\f{\pi\ell}{L}}}{n\sin{\f{\pi\ell}{n L}}} \Big)^{|\mS|}
\times  \\
\times        
 \prod^{j_1 < j_2}_{j_1,j_2 \in \mS}
         \f{\sin^2 \f{\pi (j_1 - j_2)}{n}}{\sin\f{\pi (j_1 - j_2 + \ell/L)}{n}\sin\f{\pi (j_1 - j_2 - \ell/L)}{n}},
\end{multline}
as a function of a given subset $\mS$.

For $n= n_e$ being a fixed even integer, \eqref{mFn-vertex} can be easily computed and one also can derive a compact expression for it (see \cite{work-td}).
Its analytic continuation from even integers to arbitrary real values, instead, is not simple. Remarkably, for $\Delta \alpha=1$, we were able to find the full analytic continuation
\begin{multline}
 \log \mD_{n} [1] = n \log(2\pi) - 2 \!\int_0^{\inf}\!\! \f{dt}{t}
\Big\{\f{1}{\ep^{t}-1} 
 \\
\times
\Big[
\f{(\ep^{\f{t\ell}{2L}}-1)
\big( 1 + \ep^ t \ep^{ - \f{t \ell }{2 L} } \big)}
  {2\sinh \big(\f{t\ell}{2n L}\big)}
- n \Big]
-\f{n}{2}
\Big\},\label{G3A2} 
\end{multline}
which in the replica limit simply becomes
\be
\lim_{n_e \to 1}\mD_{n_e} [1] = 2\ell /L.
\label{1anco}
\ee
The analytic continuation~\eqref{G3A2} also provides all the $n$-distances for $n$ odd, which are not given by \eqref{mFn-vertex}.

In the small interval limit, we can get further results for arbitrary $\Delta \alpha$. 
In fact, Eq.~\eqref{DrAsA} applies, e.g., for $\alpha = \alpha'$ or $\bar{\alpha} = \bar{\alpha}'$. 
In such cases the (unique) quasiprimary field satisfying \eqref{condition} is $\bar{J}$ or $J$ (respectively), with expectation 
value in the vertex operator state $|V_{\a,\bar\a}\rag$ given by $\lag \bar J \rag_{\a,\bar\a}=-2 \pi \ii \bar \a /L$ and $\lag J \rag_{\a,\bar\a}=2\pi\ii\a / L$.
 From (\ref{DrAsA}), then, we get
\bea \label{TDA1}
&& D_1(\r_{{\a,\bar\a}},\r_{{\a',\bar\a}}) = \f{|\a-\a'|\ell}{L} + o\Big(\f{\ell}{L}\Big), \nn\\
&& D_1(\r_{{\a,\bar\a}},\r_{{\a,\bar\a'}}) = \f{|\bar\a-\bar\a'|\ell}{L} + o\Big(\f{\ell}{L}\Big).
\eea
Here we used $x_J = x_{\bar J} = 1/\pi$ \cite{work-td}, as follows from (\ref{xphi}) together with the calculation of the correlation function of Refs. \cite{sierra,berganza} and its  
analytical continuation in  \cite{elc-13}.
They satisfy the bound (\ref{bound}) with $x_{\rm{max}}(1) =  1/\sr{6}$.

Another case where \eqref{condition} applies is the trace distance between vertex operators with $\bar{\alpha}=0$ or $\a=0$ and the current states $|J\rag$, $|\bar J\rag$. For such states, 
we have $\lag J \rag_J = \lag J \rag_{\bar J}=\lag \bar J \rag_J = \lag \bar J \rag_{\bar J} = 0$.
The final result in the short interval limit reads
\bea \label{TDA2}
&& D_1(\r_{J},\r_{{\a,0}}) = D_1(\r_{\bar J},\r_{{0,\a}}) = \f{|\a|\ell}{L} + o\Big(\f{\ell}{L}\Big), \nn\\
&& D_1(\r_{J},\r_{{0,\bar\a}}) = D_1(\r_{\bar J},\r_{{\bar\a,0}}) = \f{|\bar\a|\ell}{L} + o\Big(\f{\ell}{L}\Big).
\eea
The results for some other states will be given in \cite{work-td}.

\begin{figure}[t]
  \centering
  \includegraphics[width=0.45\textwidth]{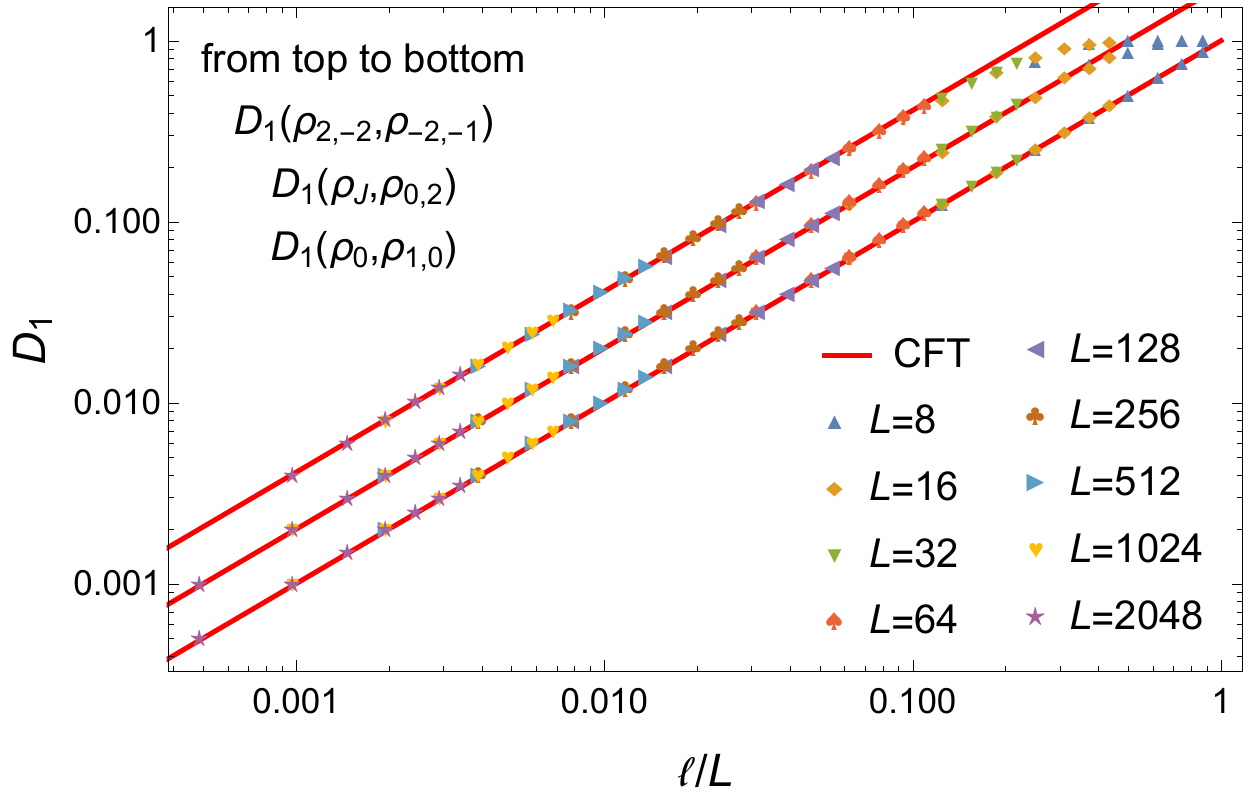}\\
  \caption{Trace distance between low-lying states as function of $\ell/L$: Numerical results for the XX spin-chain (symbols) compared with the CFT prediction (lines).
  The data are in log-log scale to show the power-law behavior at small $\ell/L$.
  Only for $\Delta \alpha=1$ (bottommost curve)  we show the analytic continuation for arbitrary $\ell/L$, cf. Eq. \eqref{1anco}, while in the other cases we compare with the 
  leading order at small $\ell/L$.}
  \label{DN}
\end{figure}

\begin{figure}[t]
  \centering
  \includegraphics[width=0.45\textwidth]{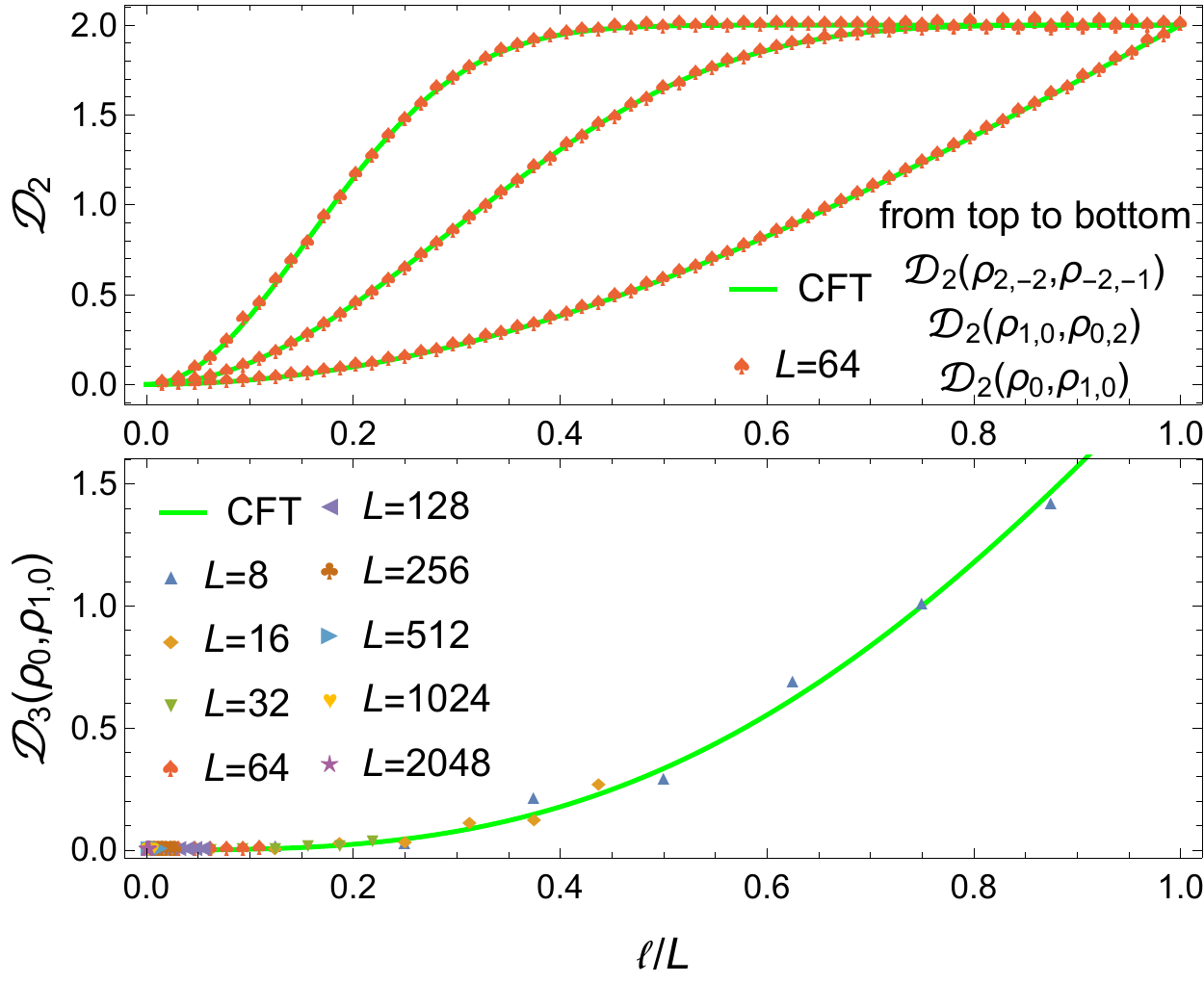}\\
  \caption{The $n$-distance for low-lying states: comparison of CFT predictions (lines) with the numerical XX spin-chain results (symbols).  
  Top: ${\cal D}_2(\r,\s)$ (CFT from Eq. \eqref{mFn-vertex}).
  Bottom: ${\cal D}_3(\r,\s)$ for $(\a-\a')^2+(\bar\a-\bar\a')^2=1$ as in Eq. \eqref{G3A2}. }
  \label{G2N}
\end{figure}

All the above results for the compact boson CFT can be checked against numerics in the XX spin chain. 
We consider a block $A$ with $\ell$ contiguous spins in a periodic chain with $L$ sites.
The correspondence between the low-lying states of the spin-chain and the CFT ones is reported, e.g., in \cite{berganza,sierra}.
As we mentioned, even for these Gaussian states  it is not possible to get trace distances in terms of the correlation functions, 
as instead one can do for the entanglement entropy \cite{vidal,peschel2003,afc-09}.
However, one can exploit the Gaussian nature of $\r_A$ and $\s_A$ \cite{peschel2003} to access the $2^{\ell}\times 2^{\ell}$ RDM numerically 
for arbitrary large $L$  \cite{vidal,peschel2003,afc-09}, 
but $\ell$ relatively small, say up to 7 (see \cite{work-td} for details). 
Only for even $n$, we can use the techniques of Refs. \cite{bb-69,fagottiXY} to obtain the $n$-distances also for very large $\ell$.
Our numerical results for many states are in Figs. \ref{DN} and \ref{G2N}.
The former shows the trace distance that asymptotically perfectly matches the CFT predictions 
(i.e. the short-interval expansion for arbitrary states and \eqref{1anco} for the vertex-states with $\Delta\alpha=1$).
Fig. \ref{G2N} reports the $2$- and $3$-distance for some states that perfectly matches available 
CFT results for arbitrary $\ell/L$. 


\paragraph{2D free massless fermion theory.}

We now consider a 2D massless free fermion theory, which is a 2D CFT with central charge $c=1/2$, and it is the continuous limit of the critical Ising spin chain.
The calculations in the 2D massless free fermion theory and Ising spin chain are parallel to those in the 2D massless free boson theory and XX spin chain.
In the CFT, besides the ground state $|0\rag$, we consider the excited states generated by the primary operators $\sigma$, $\mu$ with conformal weights $(1/16,1/16)$, $\psi$ and $\bar{\psi}$ with conformal weights $(1/2,0)$ and $(0,1/2)$ respectively, and $\ve$ with conformal weights $(1/2,1/2)$.

Making use of the results from Refs.~\cite{sierra,berganza,elc-13} we have that 
$x_\ve$ in Eq.~(\ref{xphi}) can be analytically calculated $x_\ve = 1/\pi$ \cite{work-td}.
Using Eq. (\ref{DrAsA}), and the expectation values $\lag \ve \rag_0 = \lag \ve \rag_\psi = \lag \ve \rag_{\bar \psi} = \lag \ve \rag_\ve = 0, \lag \ve \rag_\s = - \lag \ve \rag_\m = \pi /L$, we get  in the short interval limit
$D_1(\r_{0},\r_{\s}) = D_1(\r_{0},\r_{\m})=\ell/(2L)+o(\ell/L)$, 
$D_1(\r_{\s},\r_{\psi}) = D_1(\r_{\s},\r_{\bar\psi})=D_1(\r_{\m},\r_{\psi}) = D_1(\r_{\m},\r_{\bar\psi})=\ell/(2L)+o(\ell/L)$,
$D_1(\r_{\s},\r_{\ve}) = D_1(\r_{\m},\r_{\ve})=\ell/(2L)+o(\ell/L)$, and  $D_1(\r_{\s},\r_{\m})=\ell/L+o(\ell/L)$.
We checked them numerically in the critical Ising spin chain, finding perfect agreement as we will report in \cite{work-td}.

\paragraph{Conclusions.}
In this Letter we developed a replica approach to acess the trace distance in 2D QFT. 
We exploited our method to provide very general results for the distance between primary states in 2D CFT
in particular for short intervals, yielding an explicit form based on the OPE of the twist 
operators.
We also gave explicit  results for the free massless compact boson  and fermion. 
We tested analytical CFT results against the numerical trace distances in spin chains finding perfect matches.
We only reported here a subset of significant CFT results that we obtained.
We will report the calculation details and more examples  in \cite{work-td}, where we will also discuss the consequences of our findings for the 
relative entropies and fidelities. 

Our approach paves the way to systematic studies of trace distances in 2D QFT with several fundamental applications. 
First of all one can consider different geometries in CFT: open systems, disjoint intervals, finite temperature, inhomogeneous systems, etc. 
Then, one can study in a CFT after a quantum quench, how the distance between the time-dependent state and the asymptotically thermal state \cite{cc-06} 
shrinks, as well as the difference between distinct statistical ensembles. 
Massive QFT may be approached adapting the techniques of \cite{ccd-08} for the entanglement entropy.

\paragraph{Acknowledgements.}
 We thank Bin Chen, Song Cheng, and Erik Tonni for helpful discussions. All authors acknowledge support from ERC under Consolidator grant number 771536 (NEMO).
Part of this work has been carried out during the workshop  ``Entanglement in quantum systems'' at the Galileo Galilei Institute (GGI) in Florence.

\end{document}